\documentclass[10pt]{article}
\RequirePackage{fix-cm}

\title{Planar p-center problems are solvable in polynomial time when clustering a Pareto Front} 
\author{Nicolas Dupin, Frank Nielsen, El-Ghazali Talbi}

\usepackage[figuresright]{rotating}
\usepackage{amsmath,amssymb}
\usepackage{amsthm}

\usepackage{tikz,tkz-tab}

  \newtheorem{theorem}{Theorem}
  \newtheorem{lemma}{Lemma}
  \newtheorem{prop}{Proposition}
 \def\NN{{\mathbb{N}}}
 
 \def\RR{{\mathbb{R}}}
 \def\CC{{\mathcal{C}}}
 
  \def\DD{{\mathcal{D}}}
 \def\PP{{\mathcal{P}}}

\def\infty{\rotatebox{90}{8}}

\begin{document}

%
%
%
%

\maketitle

\begin{abstract}
This paper is motivated by  real-life applications of bi-objective optimization.
 Having many non dominated solutions, one wishes 
 to cluster the Pareto front using Euclidian distances.
 The p-center problems, both in the discrete and continuous versions,
are proven solvable  in polynomial time with a common dynamic   programming algorithm.
Having $N$ points to partition in $K\geqslant 3$ clusters, the complexity is proven  in $O(KN\log N)$ (resp  $O(KN\log^2 N)$) time  and $O(KN)$ memory space for the continuous (resp discrete) $K$-center problem.
$2$-center problems have  complexities in $O(N\log N)$.
To speed-up the algorithm, parallelization issues are discussed. 
A posteriori, these results allow an application
 inside multi-objective heuristics to archive partial Pareto Fronts.

 
 \end{abstract}

\section{Introduction}

This paper is motivated by real-life applications of multi-objective optimization (MOO).
Some optimization problems can be driven by more than one objective function, with some conflicts among objectives. 
For example, one may minimize financial costs, while maximizing the robustness to uncertainties \cite{dupin2015modelisation,peugeot2017mbse}.
In such cases, higher levels of robustness are likely to induce  financial over-costs.
Pareto dominance, preferring a solution from another if it is better for all the objectives, is a weak dominance rule.
With conflicting objectives, several non-dominated  solutions, denoted \emph{efficient solutions}, can be generated.
A \emph{Pareto front} (PF) is the projection in the objective space  of these efficient solutions \cite{ehrgott2003multiobjective}.

For a presentation to decision makers, concise information to summary the shape of solutions are required.
Firstly, one can present a view of a PF in clusters and the density of points in the cluster.
Secondly, representative points along the PF can be presented to support decision making.
Both problems can be seen as applications of clustering a PF, representative points being computed as the most central points of clusters.
We consider the case of two-dimensional (2d) PF using the Euclidian distance.
Having a 2d PF of size $N$, the problem is to define $K \ll N$ clusters, minimizing the dissimilarity
between the points in the clusters.
The  $K$-center problems, both in the discrete and continuous versions, define in this paper the   cluster costs, covering the 2d PF with $K$ identical balls while minimizing the radius of the balls to use. 
 The  k-center problems are NP-complete in the general case
\cite{hsu1979easy} but also for the specific case
 in $\RR^2$ using the Euclidian distance   \cite{megiddo1983new}.
This paper proves that  p-center clustering in a 2d PF is   solvable in polynomial time,
using a common Dynamic Programming (DP) algorithm, and
discuss  properties of the algorithm for an efficient implementation.

This paper is organized as following.
In Section 2, we describe the considered  problems with unified notation.
In Section 3, we discuss related state-of-the-art elements to appreciate our contributions.
 In Section 4, intermediate results are presented.
 In Section 5, a common DP algorithm is presented 
 with polynomial complexity  thanks to the results of section 4.
 In Section 6,  the implications and applications of the results of section 5 are discussed. 
 In Section 7,  our contributions  are summarized,  discussing also future directions of research.
To ease the readability of the paper, some elementary proofs are gathered in Appendix A.

\section{Problem statement and notation}

We suppose in this paper having a set $E=\{x_1,\dots, x_N\}$ of $N$ elements of $\RR^2$, 
such that for all $ i\neq j$, $x_i \phantom{0} \mathcal{I} \phantom{0} x_j$  
defining the binary relations $\mathcal{I},\prec $  for all $ y=(y^1,y^2),z=(z^1,z^2) \in \RR^2$ with:
\begin{eqnarray}
 y \prec z  & \Longleftrightarrow  & y^1< z^1 \phantom{2} \mbox{and}\phantom{2} y^2> z^2\\
y \preccurlyeq z  & \Longleftrightarrow  &  y \prec z  \phantom{2} \mbox{or}\phantom{2} y= z\\
y \phantom{1}\mathcal{I}\phantom{1} z  & \Longleftrightarrow  &y \prec z  \phantom{2} \mbox{or} \phantom{2}  z \prec y 
\end{eqnarray}

\begin{figure}[ht]
      \centering 
   \begin{tikzpicture}[scale=2.9]
    \draw[->] (-2,0) -- (2.1,0) node[right] {$x$};
    \draw[->] (0,-0.45) -- (0,0.5) node[above] {$y$};
    \draw (0.375,0) node[above] {$O=(x_O,y_O)$};
    \draw (0,0) node[color=black
    ] {$\bullet$};
    \draw (1.1,0.3) node[above] {$B=(x_B,y_B)$ with $x_O<x_B$ and $y_O<y_B$ };
    \draw (1.1,0.15) node[above] {$O$ dominates $B$};
    
    \draw (-1.1,-0.3) node[above] {$C=(x_C,y_C)$ with $x_O>x_C$ and $y_O>y_C$ };
    \draw (-1.1,-0.45) node[above] {$C$ dominates $O$};
    
    \draw (1.1,-0.3) node[above] {$D=(x_D,y_D)$ with $x_O<x_D$ and $y_D<y_O$ };
    \draw (1.1,- 0.45) node[above] {$O$ and $D$ not comparable, $O \phantom{1}\mathcal{I} \phantom{1} D$ and $O \prec D$};
    
    \draw (-1.1,0.3) node[above] {$A=(x_A,y_A)$ with $x_O>x_A$ and $y_O<y_A$ };
    \draw (-1.1,0.15) node[above] {$A$ and $O$ not comparable, $A \phantom{1} \mathcal{I} \phantom{1} O$ and $A \prec O$};

\end{tikzpicture}
    \caption{Pareto dominance and incomparability cadrans minimizing two objectives: zones of $A$ and $D$ are incomparability zones related to $O$.
    }\label{defIllustr}  
   \end{figure}
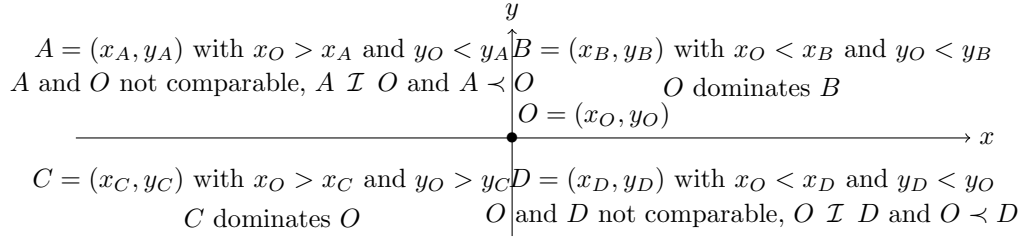

We note that the definition of $E$ 
considers minimization of two objective in the sense of inequalities defining $\mathcal{I},\prec$, as illustrated in Figure \ref{defIllustr}.
This is not a loss of generality, one can transform the objectives to maximize $f$ into $-f$ allows to consider the minimization two objectives.

\vskip 0.3cm
\noindent{
We define  $\Pi_K(E)$, as the set of all the partitions of $E$ in $K \in \NN^*$ subsets:}
\begin{equation}
\Pi_K(E) = \Big\{P \subset \PP(E)\: \Big| \:\forall p,p' \in P, \:\:p \cap p' =  \emptyset , \: \bigcup_{p \in P} p = E \: \mbox{and} \; \mbox{card}(P)=K \: \Big\} 
\end{equation}
K-center problems are combinatorial optimization problems indexed by $\Pi_K(E)$:
\begin{equation}\label{cluster}
\min_{\pi \in \Pi_K(E)}  
\max_{P \in \pi}  f(P)
\end{equation}
The cost function $f$ applied for each subset of $E$ measures a dissimilarity of the points in the subset.
In this paper, we will consider the Euclidian norm to define distances, defining for all $y=(y^1,y^2),z=(z^1,z^2) \in \RR^2$:
\begin{equation}\label{distEucl}
d(y,z) = |\!| y -z |\!| = \sqrt{ \left(y^1 - z^1\right)^2 + \left(y^2 - z^2\right)^2}
\end{equation}
The  $K$-center problems covering the PF with $K$ identical balls while minimizing the radius of the balls to use. 
The center of the balls for the discrete version are necessarily points of the 2d PF, so that 
the cost function for the discrete $K$-center problem is:
\begin{equation}\label{defKcenter}
\forall P  \subset E, \;\;\; f_{ctr}^{\DD}(P) = \min_{y \in P} \max_{x \in P}  \left|\!\left| x - y \right|\!\right|
\end{equation}
The continuous $K$-center problem is a geometric problem, minimizing the radius of covering balls without any 
 localization constraint for the center of the covering balls,
 the centers can be any subset of points in the plane:
\begin{equation}\label{defKcenterCont}
\forall P  \subset E, \;\;\; f_{ctr}^{\CC}(P) = \min_{y \in \RR^2 } \max_{x \in P}  \left|\!\left| x - y \right|\!\right|
\end{equation}
For the sake of having unified notations for common results and proofs,
we define $\gamma \in \{0,1\}$ to indicate which version of the p-center problem is considered.
$\gamma=0$ (resp $1$) indicates that the continuous (resp discrete) p-center problem is used, $f_{\gamma}$ indicates the cost measure between 
$f_{ctr}^{\CC},f_{ctr}^{\DD}$.
The continuous and discrete $K$-center problems in the 2d PF are denoted
$K$-$\gamma$-CP2dPF.

\section{Related works}

This section describes   related works to appreciate our contributions,  regarding  the 
 state of the art of the p-center problems  and clustering problems in a PF.
For more detailed surveys on the results for the p-center problems, we refer to  \cite{calik2015p}.

\subsection{Complexity results for p-center problems}

We note  that some hypotheses may be more general than the ones in Section 2.
The p-center problem consists in locating $p$ facilities among a set of possible
locations and assigning $N$ clients, called $c_1 , c_2 , \dots , c_N$ ,  to the facilities in order to minimize the maximum
distance between a client and the facility to which it is allocated.
The continuous p-center problem assumes that any place of location can be chosen, whereas the discrete
p-center problem considers a subset of $M$ potential sites denoted $f_1 , f_2 , \dots , f_M$,
and distances $d_{i,j}$ for all $i \in [\![1,n]\!]$ and $j \in [\![1,M]\!]$.
Discrete p-center problems can be formulated with bipartite graphs or infinite distances, modeling that some assignments between
clients and facilities are unfeasible.
In the discrete p-center problem defined in Section 2, as in many applications using k-center for clustering, the points $f_1 , f_2 , \dots , f_M$  are exactly $c_1 , c_2 , \dots , c_N$, 
and the distances are defined using a Euclidian norm.

Discrete and continuous p-center problems are NP-hard \cite{hsu1979easy,megiddo1984complexity}.
Furthermore, for all $\alpha<2$, any $\alpha$-approximation for the discrete p-center problem with triangle inequality  
  is NP-hard \cite{hochbaum1984np}.
\cite{hochbaum1985best,gonzalez1985clustering} provided 2-approximations for the  discrete p-center problem running respectively 
in $O(NK\log N)$ time and in $O(NK)$ time.
The  discrete p-center problem in $\RR^2$ with a Euclidian distance is also NP-hard \cite{megiddo1984complexity}.
Defining  binary variables $x_{i,j} \in \{0,1\}$ and $y_{j} \in \{0,1\}$  with 
$x_{i,j}=1$ if and only if the customer $i$ is assigned to the  depot $j$ and
$y_{j}=1$ if and only if the point $f_j$ is chosen as a depot,
the following Integer Linear Programming (ILP) formulation models the discrete p-center problem (\cite{daskin1995network}): 
\begin{equation} \label{mipPcenter}
  \begin{array}{lrlll}
 \min_{x,y,z} & z & & &\hskip 0.5cm (\ref{mipPcenter}.1)\\
  s.t: &  \sum_{j=1}^n  d_{i,j} x_{i,j} & \leqslant z&    \forall i \in [\![1,N]\!] & \hskip 0.5cm (\ref{mipPcenter}.2) \\
  & \sum_{j=1}^M y_{j} & = p & & \hskip 0.5cm (\ref{mipPcenter}.3)\\
   & \sum_{j=1}^M x_{i,j} &= 1 &   \forall i \in [\![1,N]\!] &  \hskip 0.5cm (\ref{mipPcenter}.4) \\
& x_{i,j} &\leqslant  y_{j} &   \forall (i,j) \in [\![1,N]\!] \times [\![1,M]\!], & \hskip 0.5cm (\ref{mipPcenter}.5) \\
   & x_{i,j},y_{j}& \in \{0,1\} \hskip 0.7cm&  \forall {i,j}  \in [\![1,N]\!] \times [\![1,M]\!], & \hskip 0.5cm (\ref{mipPcenter}.6)
  \end{array}
\end{equation}
Constraint (\ref{mipPcenter}.3) fixes the number of open facilities to $p$, constraints (\ref{mipPcenter}.4) assign each client to exactly one facility,
constraints (\ref{mipPcenter}.5) enforces facility $j$ to be opened to use variables $x_{i,j}$. The objective function is obtained with the linearization in 
(\ref{mipPcenter}.2).
To solve to optimality  discrete p-center problems, tighter ILP formulations, with efficient exact algorithms,
rely on the ILP models  \cite{calik2013double,elloumi2004new}.
Exponential exact algorithms were also provided for the continuous p-center problem \cite{callaghan2017speeding,drezner1984p}.
An $N^{O(\sqrt p)}$-time algorithm was provided for the  continuous Euclidean p-center problem in the plane \cite{hwang1993slab}.
An $N^{O(p^{1-1/d})}$-time algorithm for solving the continuous p-center problem in $\RR^d$ under Euclidian and $L_{\infty}$ -metric
was provided in \cite{agarwal2002exact}.
Meta-heuristics  
are also efficient for p-center problems without furnishing optimality guarantees or dual bounds \cite{ferone2017new,elshaikh2015continuous,mladenovic2003solving}.

%

Some specific cases of p-center problems are also solvable in polynomial time.
The continuous 1-center problem is exactly the  minimum covering ball problem which has a linear complexity in $\RR^2$.
Indeed,  a "prune and search" algorithm from \cite{megiddo1983linear}  finds the optimum bounding sphere and runs in linear time if the dimension is fixed as a constant. In dimension $d$, its  complexity is in { $O((d+1)(d+1)!n)$} time, which is impractical for high-dimensional applications.
The  discrete  1-center  problem,  seeking  the smallest disk centered at a point of P and containing P, can be solved solve in time  $O(N\log N)$ , using the furthest-neighbor Voronoi diagram of P \cite{brass2009computing}.
The continuous and planar  2-center problem can be solved in randomized expected $O(N \log^2 N)$ time, with algorithms provided and improved by 
\cite{sharir1997near,eppstein1997faster}.
 The discrete and planar 2-center problem  is solvable in time $O(N^{4/3}\log^5N)$ \cite{agarwal1998discrete}.
  In the case of  graded distances, the
 discrete p-center problem is polynomial time solvable when the distance matrix is graded up the rows or graded down the rows \cite{guo1998k}.
 The continuous p-center problem is  solvable in $O(N \log^3 N)$ time in a tree structure \cite{megiddo1983new}.
The  {continuous and planar  k-centers on a line},
 finding $k$ disks with centers on a given line $l$,
 is solvable in polynomial time, in $O(N^2  \log ^2 N$) time in the first algorithm by \cite{brass2009computing}
 and in $O(NK\log N)$ time and $O(N)$ space in the improved version provided by \cite{karmakar2013some}.

\subsection{Clustering/selecting points in Pareto frontiers}

Selection or clustering points in PF have been studied with applications to MOO algorithms.
Firstly, a motivation is to store representative elements of a large PF  (exponential sizes of PF are possible \cite{ehrgott2003multiobjective}) for  exact methods or population meta-heuristics.
Maximizing the quality of discrete representations of Pareto sets was studied with  the hypervolume measure 
in the Hypervolume Subset Selection (HSS) problem \cite{auger2009investigating,sayin2000measuring}. 
Secondly, a  crucial issue in the design of population meta-heuristics for MOO problems
is to select relevant solutions for operators like cross-over or mutation phases in evolutionary algorithms \cite{talbi2009metaheuristics,zio2011clustering}.
Selecting knee-points is another known approach for such goals \cite{ramirez2017knee}.


The HSS problem, maximizing the representativity of $K$ solutions among a PF of size $N$,  is known to be NP-hard in dimension 3 (and greater) since \cite{bringmann2018maximum}.
An exact  algorithm in  $n^{O(\sqrt{K})}$ and a  polynomial-time approximation scheme for any constant dimension $d$ are also provided in \cite{bringmann2018maximum}.
The 2d case is solvable in polynomial time  thanks to  a  DP algorithm 
with a complexity in $O(KN^2)$ time and $O(KN)$ space provided in \cite{auger2009investigating}.
The time complexity of the DP algorithm  was improved in $O(KN+N \log N)$ by \cite{bringmann2014two} and in $O(K.(N-K)+N \log N)$ by \cite{kuhn2016hypervolume}.

Some similar results were also proven for clustering problems.
K-median and K-medoid problems are known to be NP hard in dimension 2 since \cite{megiddo1984complexity},
the specific case of 2d PF were proven to be solvable in $O(N^3)$  time with DP algorithms \cite{dupin2018clustering,dupin2019medoids}.
K-means, one of  the  most  famous   unsupervised  learning   problem,
is also NP-hard for 2d cases \cite{mahajan2012planar}. 
The restriction to 2d PF would be also solvable in $O(N^3)$ time with a DP algorithm if a conjecture is proven \cite{dupin2018dynamic}.

We note that an affine 2d PF is a line in $\RR^2$, clustering  is equivalent to 1 dimensional cases.
 1-dimension K-means was proven to be solvable in polynomial time with a DP algorithm in $O(KN^2)$ time and $O(KN)$ space.
 This complexity was improved  for a DP algorithm in $O(KN)$ time and $O(N)$ space in \cite{gronlund2017fast}.
This is thus the complexity of K-means in an affine 2d PF.
The specific case, already mentioned in the previous section, of the continuous p-center problem with centers on a straight line
is more general that the case of an affine 2d PF, with a complexity proven in $O(NK \log N)$ time and $O(N)$ space by  \cite{karmakar2013some}.
%
 2d cases of clustering problems can also be seen as specific cases of   three-dimensional (3d) PF, affine 3d PF.
Having NP-hard complexities proven for planar cases of clustering, which is the case for k-means, p-median, k-medoids, p-center problems
since \cite{mahajan2012planar,megiddo1983new},
it implies that the considered clustering problems are also NP-hard for 3d PF.

%
%
%

\section{Intermediate results} \label{sec::singleCluster}
   
   This section gathers the intermediate result necessary for the following developments.
Some elementary proofs are gathered in Appendix A.

\subsection{Indexation  and distances in a 2d PF}
  
In this section, we analyze some properties of the relations $\prec$ and $\preccurlyeq$, before
  defining an order relation and a new indexation among the points of $E$.
  The following lemma is extends trivially  the properties of $\leqslant$ and $<$ in $\RR$:
 
\begin{lemma}
\label{transitiv}
$\preccurlyeq$ is an order relation, and 
 $\prec$ is a transitive relation:
 \begin{equation}\label{relTransitiv}
\forall x,y,z  \in \RR^2, \phantom{3} x \prec y \phantom{1} \mbox{and}\phantom{1} y \prec z \Longrightarrow x \prec z
\end{equation}
\end{lemma}

\vskip 0.3cm
The following proposition implies an order among the points of $E$, for a reindexation in $O(N\log N)$ time:

  \begin{figure}[ht]
      \centering 
   \begin{tikzpicture}[scale=0.41395]
    \draw[->] (0,0) -- (20,0) node[right] {$\emph{Obj}_1$};
    \draw[->] (0,0) -- (0,10.9) node[above] {$\emph{Obj}_2$};
    \draw (1.2,9.7) node[above] {$x_1$};
    \draw (1,9.7) node[color=blue!50] {$\bullet$};
    \draw (1.4,8.1) node[above] {$x_2$};
    \draw (1.7,8.1) node[color=blue!50] {$\bullet$};
    \draw (2.7,6.8) node[above] {$x_3$};
    \draw (2.5,6.8) node[color=blue!50] {$\bullet$};
    \draw (3.7,6.4) node[above] {$x_4$};
    \draw (3.5,6.4) node[color=blue!50] {$\bullet$};
    \draw (4.7,6.1) node[above] {$x_5$};
    \draw (4.5,6.1) node[color=blue!50] {$\bullet$};
    \draw (5.2,5.3) node[above] {$x_6$};
    \draw (5,5.3) node[color=blue!50] {$\bullet$}; 
    \draw (6.5,5.1) node[above] {$x_7$};
    \draw (6.3,5.1) node[color=blue!50] {$\bullet$};
    \draw (8.2,4.7) node[above] {$x_8$};
    \draw (8,4.7) node[color=blue!50] {$\bullet$};
    \draw (9.2,3.4) node[above] {$x_9$};
    \draw (9,3.4) node[color=blue!50] {$\bullet$};
    \draw (10.2,3.1) node[above] {$x_{10}$};
    \draw (10,3.1) node[color=blue!50] {$\bullet$};
    \draw (11.2,2.7) node[above] {$x_{11}$};
    \draw (11,2.7) node[color=blue!50] {$\bullet$};
    \draw (12.9,2.4) node[above] {$x_{12}$};
    \draw (12.7,2.4) node[color=blue!50] {$\bullet$};
    \draw (14.7,2.1) node[above] {$x_{13}$};
    \draw (14.3,2.1) node[color=blue!50] {$\bullet$};
    \draw (16.6,1.4) node[above] {$x_{14}$};
    \draw (16.4,1.4) node[color=blue!50] {$\bullet$};
    \draw (18.9,0.9) node[above] {$x_{15}$};
    \draw (18.7,0.9) node[color=blue!50] {$\bullet$};
    \end{tikzpicture}
    \caption{Illustration of a 2d PF with 15 points and the indexation implied by Proposition \ref{reord} }\label{orderIllustr}  
   \end{figure}
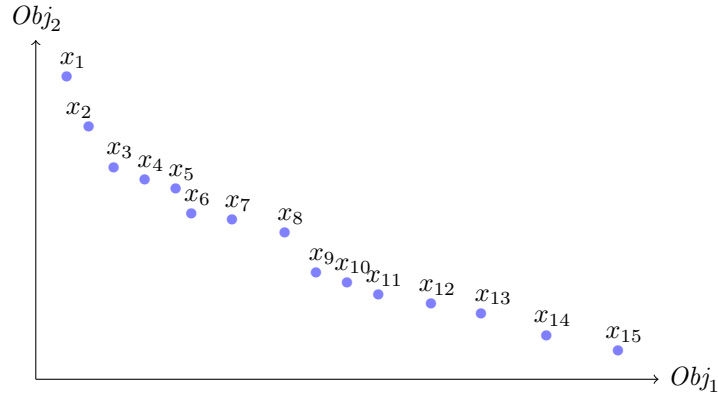

\begin{prop}[Total order]\label{reord}
Points $(x_i)$ can be indexed such that:
\begin{eqnarray}
\forall (i_1,i_2) \in [\![1;N]\!]^2, \phantom{3} & i_1<i_2 \Longrightarrow & x_{i_1} \prec x_{i_2}\label{ordCroissant} \\
\forall (i_1,i_2) \in [\![1;N]\!]^2, \phantom{3} &  i_1\leqslant i_2 \Longrightarrow & x_{i_1} \preccurlyeq x_{i_2} \label{ordCroissant2}
\end{eqnarray}
This property is stronger than the property that $\preccurlyeq$ induces a total order in $E$.
Furthermore, the complexity of the sorting re-indexation is in $O(N\log N)$
\end{prop}

\noindent{\textbf{Proof}}:
We index $E$ such that the first coordinate is increasing:
$$\forall (i_1,i_2) \in [\![1;N]\!]^2,  i_1<i_2 \Longrightarrow  x_{i_1}^1 < x_{i_2}^1$$
This sorting procedure has a complexity in $O(N.\log N)$.
Let $(i_1,i_2) \in [\![1;N]\!]^2$, with $i_1<i_2$. We have thus $x_{i_1}^1 < x_{i_2}^1$.
Having $x_{i_1} \mathcal{I} x_{i_2}$ implies $x_{i_1}^2 > x_{i_2}^2$.
$x_{i_1}^1 < x_{i_2}^1$ and $x_{i_1}^2 > x_{i_2}^2$ is by definition $x_{i_1} \prec x_{i_2}$. $\square$ 

\vskip 0.3cm
The re-indexation of Proposition \ref{reord} implies also a monotonic structure for the 
distances among points of the 2d PF:
\begin{lemma}\label{orderDist}
We suppose that points $(x_i)$ are sorted following Proposition \ref{reord}.
\begin{eqnarray}
\forall (i_1,i_2,i_3) \in [\![1;N]\!]^3,&   i_1 \leqslant i_2<i_3 \Longrightarrow d(x_{i_1},x_{i_2}) < d(x_{i_1},x_{i_3}) \label{eq1} \\
\forall (i_1,i_2,i_3) \in [\![1;N]\!]^3, &   i_1<i_2 \leqslant i_3 \Longrightarrow d(x_{i_2},x_{i_3}) < d(x_{i_1},x_{i_3}) \label{eq2}
\end{eqnarray}
\end{lemma}

\subsection{1-center in a 2d PF}

 1-center optimization problems have a trivial solution,
the unique partition of $E$ in one subset is $E$.
To solve 1-center problems, it is required to compute the cost of the trivial partition,
i.e. to compute the radius of minimum enclosing disk covering all the points of E (and centered in one point of $E$ for the discrete version).
This section provides the complexity result related to these problems.

\begin{lemma}\label{costOptPcenterCont}
Let $P \subset E$ such that $\mbox{card} (P)\geqslant 1$. 
Let $i$ (resp $i'$) the minimal (resp maximal) index of points of $P$.
%
\begin{equation}\label{costOptPcenterCont::eq}
 f_{ctr}^{\CC}(P) =   \frac 1 2 \left|\!\left| x_i - x_{i'} \right|\!\right|
\end{equation}
\end{lemma}

%

\begin{lemma}\label{calcKcenter}
Let $P \subset E$ such that $\mbox{card} (P)\geqslant 1$. 
Let $i$ (resp $i'$) the minimal (resp maximal) index of points of $P$.
 \begin{equation}\label{costOptPcenterDiscr::eq}
f_{ctr}^{\DD}(P) 
=  \min_{j \in [\![i,i']\!], x_j \in P} \max \left(\left|\!\left| x_j - x_i \right|\!\right|,\left|\!\left| x_j - x_{i'} \right|\!\right|\right)
\end{equation}

\end{lemma}



\begin{prop}\label{1center}
Let $\gamma \in \{0,1\}$, let  $E =\{x_1,\dots, x_N\}$ a subset of $N$ points  of $\RR^2$, 
such that for all $ i\neq j$, $x_i \phantom{0} \mathcal{I} \phantom{0} x_j$.
$1$-$\gamma$-CP2dPF has a complexity in $O(N)$ time.
\end{prop}

\noindent{\textbf{Proof}}: Using equations (\ref{costOptPcenterCont::eq}) or (\ref{costOptPcenterDiscr::eq}), computations of $f_{\gamma}$
is in $O(N)$ once having computed the extreme elements following the order $\prec$.
Computing the extreme points is also in $O(N)$, with one traversal of the elements of $E$.
Finally, the complexity of 1-center problems are in linear time. $\square$

\subsection{Optimality of interval clustering}

In this section, a common characterization of optimal solutions of problems (\ref{cluster}) for  the p-center problems is given in Proposition
\ref{propOptcPcent}.
The proof is common for the discrete and continuous cases, using Lemma
\ref{inclusionLemma}. 

\begin{lemma}\label{inclusionLemma}
Let $\gamma \in \{0,1\}$. Let $P \subset P' \subset E$. 
We have $f_{\gamma}(P) \leqslant f_{\gamma}(P') $.
\end{lemma}

%
%
%

\begin{prop}
\label{propOptcPcent}\label{propOptdPcent}
Let $\gamma \in \{0,1\}$, let $K \in \NN$, let $E=(x_i)$ a 2d PF,
indexed following Proposition \ref{reord}.
There exists optimal solutions of $K$-$\gamma$-CP2dPF 
 using only clusters  $\CC_{i,i'} = \{x_j\}_{j \in [\![i,i']\!]}= \{x \in E \: | \: \exists j \in [\![i,i']\!],\: x = x_j \} $.
\end{prop}

\noindent{\textbf{Proof}:}  We prove the result by induction on $K \in \NN$.
For $K=1$, the optimal solution is unique, the optimal cluster is $E = \{x_j\}_{j \in [\![1,N]\!]}$.
Let us suppose $K>1$ and the Induction Hypothesis (IH) that Proposition \ref{propOptcPcent}  is true for $K$-$\gamma$-CP2dPF.
Let $\pi \in \Pi_K(E)$ an optimal solution of $K$-$\gamma$-CP2dPF, let $OPT$ the optimal cost.

We denote $\pi = \CC_1, \dots, \CC_K$ the $K$ subsets of the partition $\pi$ denoting $\CC_K$ the cluster of $x_N$.
For all $k \in [\![1,K]\!]$, $f_{\gamma}(\CC_k) \leqslant OPT$.
Let $i$ the minimal index such that $x_i \in \CC_K$.
 We consider the subsets
$\CC_K' = \{x_j\}_{j \in [\![i,N]\!]}$ and 
$\CC_k' = \CC_k \cap \{x_j\}_{j \in [\![1,i-1]\!]}$ for all $k \in [\![1,K-1]\!]$.
With such construction, it is clear that $\CC_1', \dots, \CC_{K-1}'$ is a partition of $\{x_j\}_{j \in [\![1,i-1]\!]}$,
and $\CC_1', \dots, \CC_K'$ is a partition of $E$.
For all $k \in [\![1,K]\!]$, $\CC_k' \subset \CC_k$ so that $f_{\gamma}(\CC_k') \subset f_{\gamma}(\CC_k)  \leqslant OPT$
(Lemma \ref{inclusionLemma}).

$\CC_1', \dots, \CC_K'$ is a partition of $E$, and $\max_{k \in [\![1,K]\!]} f(\CC_k)  \leqslant OPT$.
$\CC_1', \dots, \CC_K'$ is an optimal solution of $K$-$\gamma$-CP2dPF.
$\CC_1', \dots, \CC_{K-1}'$ is an optimal solution of the considered $K-1$-$\gamma$-CP2dPF
applied to points $E' = \cup_{k=1}^{K-1} \CC_1'$.
Let $OPT'$ the optimal cost, we have $OPT' \leqslant  \max_{k\in [\![1,K-1]\!]}f_{\gamma}(\CC_k') \leqslant  OPT$.
Applying  IH for $K-1$-$\gamma$-CP2dPF to  points $E'$,
we have $\CC_1'', \dots, \CC_{K-1}''$ an optimal solution of $K-1$-$\gamma$-CP2dPF among $E'$ on the shape
$\CC_{i,i'} = \{x_j\}_{j \in [\![i,i']\!]}= \{x \in E' \: | \: \exists j \in [\![i,i']\!],\: x = x_j \} $.
For all $k \in [\![1,K-1]\!]$, $f_{\gamma}(\CC_k'') \leqslant OPT' \leqslant OPT$.
$\CC_1'', \dots, \CC_{K-1}'', \CC_K'$ is finally an optimal solution of $K$-$\gamma$-CP2dPF in $E$
using only clusters on the shape $\CC_{i,i'}$. $\square$

\begin{figure}[ht]
 \centering 
\begin{tabular}{ l }
\hline
\textbf{Algorithm 1: Computation of $ f_{ctr}^{\DD}(\CC_{i,i'})$}\\
\hline
\verb!  !  \textbf{input}: indexes $i<i'$\\
\verb!  !  \textbf{output}: the cost  $f_{ctr}^{\DD}(\CC_{i,i'})$ 
\\ 
\textbf{Initialization:}\\ 
\verb!  ! define $\mbox{idInf}=i$, $\mbox{valInf}=\left|\!\left| x_i - x_{i'} \right|\!\right|$, \\
\verb!  ! define $\mbox{idSup}=i'$, $\mbox{valSup}=\left|\!\left| x_i - x_{i'} \right|\!\right|$, \\
\verb!  ! \textbf{while} $\mbox{idSup}-\mbox{idInf}\geqslant 2$ \verb!  ! //Dichotomic search\\
\verb!    ! Compute $\mbox{idMid}= \left \lfloor \frac {i+i'} 2 \right \rfloor$, $\mbox{valTemp}= f_{i,i',\emph{idMid}}$, $\mbox{valTemp2}= f_{i,i',\emph{idMid}+1}$\\
\verb!    ! \textbf{if} $\mbox{valTemp}=\mbox{valTemp2}$ \\
\verb!      !$\mbox{idInf}=\mbox{idMid}, \mbox{valInf}=\mbox{valTemp}$ \\
\verb!      !$\mbox{idSup}=1+\mbox{idMid}$, $\mbox{valSup}=\mbox{valTemp2}$ \\
\verb!    ! \textbf{if} $\mbox{valTemp}<\mbox{valTemp2}$ // increasing phase\\
\verb!      !$\mbox{idSup}=\mbox{idMid}, \mbox{valSup}=\mbox{valTemp}$ \\
\verb!    ! \textbf{if} $\mbox{valTemp}>\mbox{valTemp2}$ \\
\verb!      !$\mbox{idInf}=1+\mbox{idMid}, \mbox{valInf}=\mbox{valTemp2}$ \\

\verb!  ! \textbf{end while} \\
\textbf{return} $\min (\mbox{valInf},\mbox{valSup})$ \\
\hline
\end{tabular}
\end{figure}

 \subsection{Computation of cluster costs}

Proposition \ref{propOptcPcent} induces that only costs of  clusters $\CC_{i,i'}$ shall be computed.
The issue is here to compute the most efficiently such clusters.
Once points are sorted following Proposition 1, equation (\ref{costOptPcenterCont::eq})
assures that cluster costs  $f_{ctr}^{\CC}(\CC_{i,i'})$ can be computed in $O(1)$ once $E$ is sorted following Proposition \ref{reord}. 


Equation (\ref{costOptPcenterDiscr::eq}) assures 
that cluster costs  $f_{ctr}^{\DD}(\CC_{i,i'})$ can be computed in $O(i'- i)$ for all $i<i'$.
Actually, Lemma \ref{monotony} and Proposition \ref{propCostPcentDiscr} allows to have computations in $O(\log(i'- i))$
once points are sorted following Proposition 1. It allows to compute all the  
$f_{ctr}^{\DD}(\CC_{i,i'})$ for all $i<i'$ in $O(N^2 \log N)$.

\begin{lemma}\label{monotony}
Let $(i,i')$ with $i<i'$.
$f_{i,i'} : j  \in [\![i,i']\!] \longmapsto  
\max \left(\left|\!\left| x_j - x_i \right|\!\right|,\left|\!\left| x_j - x_{i'} \right|\!\right|\right)$ is strictly decreasing before reaching first a minimum $f_{i,i'}(l)$, 
$f_{i,i'}(l+1) \geqslant f_{i,i'}(l)$, and then is strictly increasing for $j  \in [\![l+1,i']\!]$
\end{lemma}


\begin{prop}\label{propCostPcentDiscr},
Let  $E =\{x_1,\dots, x_N\}$ be $N$ points  of $\RR^2$, 
such that for all $ i\neq j$, $x_i \prec x_j$.
Computing cost $f_{ctr}^{\DD}(\CC_{i,i'})$  for any  cluster
$\CC_{i,i'}$
has a complexity in $O(\log (i'-i))$ time.
\end{prop}

\noindent{\textbf{Proof}:} 
Let $i<i'$. Algorithm 1 uses Lemma \ref{monotony} to have as a loop invariant
the existence of a minimal solution of $f_{i,i'}(j*)$ with  $\mbox{idInf} \leqslant j* \leqslant \mbox{idSup}$.
%
Indeed, Algorithm 2 computes $l= \mbox{idMid}= \left \lfloor \frac {\mbox{idInf}+\mbox{idSup}} 2 \right \rfloor$ and the values $ f_{i,i'}(l)$ and  $f_{i,i'}(l+1)$, operations in $O(1)$.
If $ f_{i,i'}(l) <f_{i,i'}(l+1)$, the Lemma \ref{monotony} ensures that the center of $\CC_{i,i'}$ is before $l$; so that it can be reactualized $\mbox{idSup}=l$.
If $ f_{i,i'}(l) >f_{i,i'}(l+1)$, the Lemma \ref{monotony} ensures that the center of $\CC_{i,i'}$ is after $l+1$; so that it can be reactualized $\mbox{idInf}=l+1$.
 Iterating this procedure till $M-m<2$ finds the center of  $\CC_{i,i'}$ using at most $\log(i'-i)$ operations in $O(1)$. 
 $\square$

\section{DP algorithm and complexity results}

 Proposition \ref{propOptcPcent}  implies that 
optimal solutions of  the p-center problems (\ref{cluster}) 
can be designed using only subsets $\CC_{i,i'}$.
Enumerating such partitions with a brute force algorithm
would lead to $\Theta(N^K)$ computations of the cost of the partition.
This is not enough to guarantee to have a polynomial algorithm,
but it is a first step  for the clustering algorithm of section 5.

\begin{figure}[ht]
 \centering 
\begin{tabular}{ l }
\hline
\textbf{Algorithm 2: p-center clustering in a 2dPF, general DP algorithm}\\
\hline

\textbf{Input:} \\
- $N$ points  of $\RR^2$, $E =\{x_1,\dots, x_N\}$  
such that for all $ i\neq j$, $x_i \phantom{0} \mathcal{I} \phantom{0} x_j$ ;\\
- $\gamma \in \{0,1\}$ to specify the clustering measure;\\ 
- $K\in\NN$ the number of clusters.\\
\textbf{Output:}  $OPT$ the optimal cost  and the clustering partition $\PP$\\

\verb!  ! initialize  matrix $C$ with  $C_{i,k}=0$  for all $i\in [\![1;N]\!], k\in [\![1;K-1]\!]$\\
\verb!  ! sort $E$ following the order of Proposition \ref{reord}\\
\verb!  ! compute $C_{i,1} = f_{\gamma}(\CC_{1,i})$ for all $i\in [\![1;N]\!]$\\
\verb!  ! \textbf{for} $k=2$ to $K-1$ \verb!  !//Construction of the matrix $C$ \\
\verb!    ! \textbf{for} $i=2$ to $N-1$ \\
\verb!      ! set $C_{i,k} = \min_{j \in [\![2,i]\!]} \max(C_{j-1,k-1} ,f_{\gamma}(\CC_{j,i}))$\\
\verb!    ! \textbf{end for} \\
\verb!  ! \textbf{end for} \\
\verb!  ! set $OPT = \min_{j \in [\![2,N]\!]} \max(C_{j-1,N-1} ,f_{\gamma}(\CC_{j,N}))$\\
\verb!  ! set $j = \mbox{argmin}_{j \in [\![2,N]\!]} \max(C_{j-1,N-1} ,f_{\gamma}(\CC_{j,N}))$\\

\verb!  ! $i=j$ \verb!  !//Backtrack phase\\
\verb!  ! initialize $\PP=\{[\![j;N]\!]\}$, a set of sub-intervals of $[\![1;N]\!]$.\\
\verb!  ! \textbf{for} $k=K$ to $1$ with increment $k \leftarrow k-1$\\
\verb!    ! find $j\in [\![1,i]\!]$ such that $\min_{j \in [\![1,i]\!]} \max(C_{j-1,k-1} ,f_{\gamma}(\CC_{j,i}))$\\
\verb!    ! add $[\![j,i]\!]$ in $\PP$\\
\verb!    ! 
$i=j-1$\\
\verb!  ! \textbf{end for} \\

\textbf{return}  $OPT$ the optimal cost  and the partition $\PP$\\
\hline
\end{tabular}
\end{figure}

\subsection{General DP algorithm }

 Proposition  \ref{propOptcPcent}  allows to derive a common DP algorithm for p-center problems. 
 Defining $C_{i,k}$ as the optimal cost of $k$-$\gamma$-CP2dPF  clustering with $k$ cluster among points $[\![1,i]\!]$ for all $i \in [\![1,N]\!]$ and $k \in [\![1,K]\!]$.
 The case $k=1$ is given by:
\begin{equation}
\forall i \in [\![1,N]\!], \:\:\: C_{i,1} =  f_{\gamma}(\CC_{1,i})
\end{equation}
We have following induction relation, with the convention  $C_{0,k} = 0$ for all $k\geqslant0$:
\begin{equation}\label{inducForm}
\forall i \in [\![1,N]\!], \: \forall k \in [\![2,K]\!], \:\:\: C_{i,k} = \min_{j \in [\![1,i]\!]}  \max(C_{j-1,k-1}, f_{\gamma}(\CC_{j,i}))
\end{equation}
Algorithm 2 uses these relations to compute the optimal values of $C_{i,k}$.
$C_{N,K}$ is the optimal solution of $K$-$\gamma$-CP2dPF, a backtracking algorithm allows to compute the optimal partitions.

\subsection{Computing the lines of the DP matrix}

In Algorithm 2, the main issue to calculate the complexity is 
to compute efficiently  $C_{i,k} = \min_{j \in [\![2,i]\!]} \max(C_{j-1,k-1} ,f_{\gamma}(\CC_{j,i}))$.
Lemma \ref{monotony2} and Proposition \ref{complexLineDP}
allows to compute each line of the DP matrix with  a time complexity in $O(N \log N)$ and $O(N)$ space.

\begin{lemma}\label{monotony02}
Let  $k \in  [\![2,K]\!]$.
The application $g_{i,k} : j  \in [\![1,N]\!] \longmapsto  C_{j,k}$ is  increasing. 
\end{lemma}

\begin{lemma}\label{monotony2}
Let  $i \in  [\![2,N]\!]$, $k \in  [\![2,K]\!]$.
Let $g_{i,k} : j  \in [\![2,i]\!] \longmapsto  \max(C_{j-1,k-1} ,f_{\gamma}(\CC_{j,i}))$.
It exists $l  \in [\![2,i]\!]$ such that $g_{i,k}$ is  decreasing for $j  \in [\![2,l]\!]$, and then is  increasing for $j  \in [\![l+1,i]\!]$
\end{lemma}

\begin{figure}[ht]
 \centering 
\begin{tabular}{ l }
\hline
\textbf{Algorithm 3: Dichotomic computation of $C_{i,k} = \min_{j \in [\![2,i]\!]} \max(C_{j-1,k-1} ,f_{\gamma}(\CC_{j,i}))$}\\
\hline
\verb!  !  \textbf{input}: indexes $i \in  [\![2,N]\!]$, $k \in  [\![2,K]\!]$,  $\gamma \in \{0,1\}$,
a vector $v_j =C_{j,k-1}$ for all $j \in [\![1,i-1]\!]$.\\
\verb!  !  \textbf{output}: $\min_{j \in [\![2,i]\!]} \max(C_{j-1,k-1} ,f(\CC_{j,i}) )$ 
\\ 
\textbf{Initialization:}\\ 
\verb!  ! define $\mbox{idInf}=2$, $\mbox{valInf}=f_{\gamma}(\CC_{2,i}) $, \\
\verb!  ! define $\mbox{idSup}=i$, $\mbox{valSup}=v_{i-1} $, \\
\verb!  ! \textbf{while} $\mbox{idSup}-\mbox{idInf}>2$ \verb!  ! //Dichotomic search\\
\verb!    ! Compute $\mbox{idMid}= \left \lfloor \frac {i+i'} 2 \right \rfloor, \mbox{valTemp}= g_{i,k}(\emph{idMid}), \mbox{valTemp2}= g_{i,k}(\emph{idMid}+1)$\\
\verb!    ! \textbf{if} $\mbox{valTemp}<\mbox{valTemp2}$ // increasing phase\\
\verb!      !$\mbox{idSup}=\mbox{idMid}, \mbox{valSup}=\mbox{valTemp}$ \\
\verb!    ! \textbf{else} \\
\verb!      !$\mbox{idInf}=1+\mbox{idMid}, \mbox{valInf}=\mbox{valTemp2}$ \\

\verb!  ! \textbf{end while} \\
\textbf{return} $\min (\mbox{valInf},\mbox{valSup})$ \\
\hline
\end{tabular}
\end{figure}

\begin{prop}[Line computation]\label{complexLineDP}
Let $i \in  [\![2,N]\!], k \in  [\![2,K]\!]$. Let $\gamma \in \{0,1\}$.
Once the values $C_{j,k-1}$ in the DP matrix of Algorithm 2 are computed,
Algorithm 3 computes $C_{i,k} = \min_{j \in [\![2,i]\!]} \max(C_{j-1,k-1},f_{\gamma}(\CC_{j,i}))$ calling $O(\log i)$ cost computations $f_{\gamma}(\CC_{j,i})$.
It induces a time complexity in $O(\log^{\gamma} i)$.
In other words, once the line  of the DP matrix $C_{j,k-1}$ is computed for all $j \in [\![1,N]\!]$, 
the line  $C_{j,k-1}$ can be computed
with  a complexity in $O(N \log^{1+\gamma} N)$  time and $O(N)$ space.
\end{prop}
\noindent{\textbf{Proof}:} Algorithm 3 is a dichotomic (and thus logarithmic) search based on Lemma \ref{monotony2}, similarly to Algorithm 1 derived from Lemma 
\ref{monotony}.
The complexity to call Algorithm 3 is $O(\log i)$ cost computations $f_{\gamma}(\CC_{j,i})$. $\square$

%
%

\subsection{Linear memory consumption}

At this stage, the time complexity of Algorithm 1 can be provided
with a space complexity in $O(KN)$, the size of the DP matrix.
This section aims to reduce this complexity into a $O(N)$ memory space.

Actually, the DP matrix $C$ can be computed line by line, with the index $k$ increasing.
The computation of line $k+1$ requires only the line $k$ and computations of cluster costs requiring $O(1)$ additional memory space.
In the DP matrix, deleting the line $k-1$ once the line $k$ is completed
allows to have $2N$ elements in the memory?
IT allows to compute the optimal value $C_{N,K}$ using $O(N)$ memory space.
The point here is that the backtracking operations, as written in Algorithm 2, require stored values of the whole matrix.
This section aims to provide an alternative backtrack algorithm
with a complexity in at most $O(N)$ memory space and $O(KN\log N)$ time.

We define Algorithm 4,an algorithm starting from the point $z_N$,
 computing the last cluster are the largest one with a size below the optimal
cost of p-center an,d iterating successivelmy by constructing the largest cluster.
We prove that this procuedure furnish optimal partitions for the p-center problems.

\begin{figure}[ht]
 \centering 
\begin{tabular}{ l }
\hline
\textbf{Algorithm 4: Backtracking algorithm using $O(N)$ memory space}\\
\hline
\verb!  !  \textbf{input}: - $\gamma \in \{0,1\}$ to specify the clustering measure;\\
\verb!    !- $N$ points  of a 2d PF, $E =\{z_1,\dots, z_N\}$, sorted 
such that for all $ i< j$, $z_i \phantom{0} \prec \phantom{0} z_j$ ;\\
\verb!    !- $K\in\NN$ the number of clusters;\\
\verb!    !- $OPT$, the optimal cost of $K$-$\gamma$-CP2dPF;\\

\verb!  !  \textbf{output}: $\PP$ an optimal partition of $K$-$\gamma$-CP2dPF.\\
\verb!  !\\ 
\verb!  ! initialize $maxId = N$, $minId = N$, $\PP=\varnothing$, a set of sub-intervals of $[\![1;N]\!]$.\\
\verb!  ! \textbf{for} $k=K$ to $2$ with increment $k \leftarrow k-1$\\
\verb!    ! set $minId = maxId$\\
\verb!    ! \textbf{while} $f_{\gamma}(\CC_{minId-1,maxId})) \leqslant OPT$ \textbf{do} $minId=minId$ \textbf{end while}\\
\verb!    ! add $[\![minId,maxId]\!]$ in $\PP$\\
\verb!    ! set $maxId = minId-1$\\
\verb!  ! \textbf{end for} \\
\verb!  ! add $[\![1,maxId]\!]$ in $\PP$\\
\textbf{return} $\PP$ \\
\hline
\end{tabular}
\end{figure}

\begin{lemma}\label{backtrackOrderLemma}
Let $K \in \NN, K\geqslant 2$. 
Let  $E =\{z_1,\dots, z_N\}$, sorted 
such that for all $ i< j$, $z_i \phantom{0} \prec \phantom{0} z_j$.
For the discrete and continuous $K$-center problems,
the indexes given by Algorithm 4 are lower bounds
of the indexes of any optimal solution:
Denoting $[\![1,i_1]\!],[\![i_1+1,i_2]\!], \dots, [\![i_{K-1}+1,N]\!]$
the indexes given by Algorithm 4,
and $[\![1,i'_1]\!],[\![i'_1+1,i'_2]\!], \dots, [\![i'_{K-1}+1,N]\!]$
the indexes of an optimal solution,
we have for all $k \in [\![1,K-1]\!], i_k \leqslant i'_k$
\end{lemma}

\begin{prop}\label{backtrackDP}
Once the  optimal cost of p-center problems are computed,
Algorithm 4 computes an optimal partition 
in $O(N \log N)$ time using $O(1)$ additional memory space.
\end{prop}
\noindent{\textbf{Proof}:} 
Let  $OPT$, the optimal cost of $K$-center clustering with $f$.
Let $[\![1,i_1]\!],[\![i_1+1,i_2]\!], \dots, [\![i_{K-1}+1,N]\!]$
the indexes given by Algorithm 4.
By construction, all the clusters $\CC$ defined by the indexes $[\![i_k+1,i_{k+1}]\!]$ for all $k>1$ verify $f_{\gamma}(\CC) \leqslant OPT$.
Let $\CC_1$ the cluster defined by $[\![1,i_1]\!]$,
we have to prove that $f_{\gamma}(\CC_1) \leqslant OPT$ to conclude of the optimality
of the clustering defined by Algorithm 4.
Let an optimal solution, let $[\![1,i'_1]\!],[\![i'_1+1,i'_2]\!], \dots, [\![i'_{K-1}+1,N]\!]$ the indexes defining this solution.
Lemma \ref{backtrackOrderLemma} ensures that $i_1 \leqslant i'_1$, and thus 
Lemma \ref{inclusionLemma} assures $f_{\gamma}(\CC_{1,i_1}) \leqslant f_{\gamma}(\CC_{1,i'_1}) \leqslant OPT$.

Analyzing the complexity, Algorithm 4 calls at most
$(K+N)\leqslant 2N$ times the clustering cost function, without 
requiring stored elements,
the complexity is in $O(N \log^{{\gamma}} N)$ time. $\square$
\vskip 0.3cm
\noindent{\textbf{Remark}  Actually, finding the biggest cluster with an extremity given and a bounded cost
 can be proceeded by a dichotomic search. It would induce a complexity in $O(K \log^{{1+\gamma}} N)$.
 To avoid the separate case $K=O(N)$ and $\gamma=1$, Algorithm 4 provides
 a common algorithm running in $O(N \log N)$ time which is enough for the following complexity results.}

\subsection{Complexity results}

\begin{theorem}[Polynomial complexities for K-centers in a 2d PF]
Let  $E =\{x_1,\dots, x_N\}$ a subset of $N$ points  of $\RR^2$, 
such that for all $ i\neq j$, $x_i \phantom{0} \mathcal{I} \phantom{0} x_j$.
1-center problems are solvable in linear time.
Applied to the 2d PF $E$ for $p\geqslant 2$, the p-center problems are solvable to optimality in polynomial time using Algorithm 2, and its operators defined in Algorithms 3 and 4.
For $K\geqslant 2$, the continuous K-center problems are solvable in $O(KN\log N)$ time and $O(N)$ space.
The discrete 2-center is solvable in $O(N\log N)$ time and $O(N)$ space.
For $K\geqslant 3$, the  discrete K-center problems are solvable in $O(KN\log^2 N)$ time and $O(N)$ space.
\end{theorem}

\noindent{\textbf{Proof}:}
The induction formula (\ref{inducForm}) uses only values $C_{i,j}$ with $j<k$ in Algorithm 3.
$C_{N,k}$ is  at the end of each loop in $k$ the optimal value
of the $k$-center clustering among the $N$ points of $E$,
and the optimal cost.
Proposition \ref{backtrackDP} ensures that Algorithm 4
gives an optimal partition, which proves the validity of Algorithm 1.

Let us analyze the complexity. We suppose $K\geqslant 2$, the case $K=1$ is given by Proposition \ref{1center}.
The space complexity is in $O(N)$ using section 5.3.
Sorting and indexing the elements of $E$ following Proposition \ref{reord} has a time complexity in $O(N\log N)$.
Computing the first line of the DP matrix costs has also a time complexity in $O(N\log N)$
With Proposition \ref{complexLineDP}
The construction of the lines of the DP matrix $C_{i,k}$ for $k \in [\![2,K-1]\!]$  requires $N\times (K-2)$
computations of $\min_{j \in [\![1,i]\!]} C_{j-1,k-1} + f_{\gamma}(\CC_{j,i})$, which are in $O(\log^{1+\gamma} N)$ time,
the complexity of this phase is in  $O((K-2)N \log^{1+\gamma} N)$ .
Proposition \ref{backtrackDP} ensures that backtracking operations are also in $O(N\log N)$ time.
Finally, the final time complexities are in $O(N\log N + (K-2)N \log^2 N)$
for the discrete $K$-center problem, and in $O(N\log N + (K-2)N \log N) = O(KN\log N) $ for the continuous $K$-center problem.
For the discrete p-center problem, the case $K=2$ induces a specific complexity in $O(N\log N)$, whereas cases $K \geqslant 3$ imply a time complexity in
$O(KN\log^2 N)$. $\square$


\section{Discussions}


\subsection{Importance of the 2d PF hypothesis}

Planar p-center problems were not studied previously in the case of a PF.
This hypothesis, leading to many applications in bi-objective optimization,
is crucial for the complexity results.
The p-center problems were proven NP-hard in a planar Euclidean space  since \cite{megiddo1984complexity}. Adding the PF hypothesis induces  the polynomial complexity of Theorem 1 and allows an efficient implementation of the algorithm. Two properties of 2d PF were crucial for these results. On one hand, the  1-dimensional structure implied by  Proposition \ref{reord}  allows an extension of DP algorithms \cite{wang2011ckmeans,gronlund2017fast}.
On the other hand, properties (\ref{costOptPcenterCont::eq}) and (\ref{costOptPcenterDiscr::eq})
allow fast computations of cluster costs.
The specific computations of the minimum enclosing disks
allows to improve in the 2d PF case the general results for planar Euclidian spaces. 
The discrete 1-center problem is proven $O(N )$ time for 2d PF,
instead of $O(N \log N)$ (\cite{brass2009computing}).
The  continuous 1-center problem has a complexity in $O(N)$ time proven,
as in  in \cite{megiddo1983linear}, but with a much simpler algorithm.
Furthermore, when having to compute many cluster costs like in Algorithm 2,
the initial sorting following Proposition \ref{reord} is amortized,
with marginal computations to compute cluster costs in $O(N \log^{\gamma} N)$ time.
The complexity continuous 2-centre case is also improved, in $O(N \log N)$ time
instead of $O(N \log^2 N)$ time in \cite{eppstein1997faster,sharir1997near}.

\subsection{Equivalent optimal solutions}
 
 Section 5.3 emphasizes that many optimal solution may exist.
 The backtracking procedure in Algorithm 4 gives the clusters
 with the minimal indexes.
An alternative backtracking procedure, constructing the first clusters starting from a point $x_i$, initially $x_1$, and furnishing the maximal clusters represented by 
 $[\![i,j]\!]$ of cost bounded by $OPT$, furnishes also optimal solutions with the maximal indexes.
 A large funnel of optimal solutions may exist.
Many optimal solutions may be nested, i.e. non verifying the Proposition  \ref{propOptdPcent}.
 For the real-life application, having well-balanced clusters is more natural. The Algorithm 4 provides badly balanced solutions.
 One may will to balance the sizes of covering balls, or the number of points in the clusters.
 Both types of solutions may be given using simple and fast post-processing.
 For example, one may proceed a steepest descent local search using 2-center problem types for consecutive clusters in the current solution.
 For balancing the size of clusters, iterating 2-center computations
 induces marginal computations in
 $O(\log^{1+\gamma} N)$ time for each iteration.

 \subsection{Variants of the problem}
 
 The results  can be extended to some variants of the p-center problems.
The rectilinear p-center considers the $L_{\infty}$ norm \cite{drezner1987rectangular}.
Considering rectilinear p-center problems,
Proposition \ref{orderDist} is also true, and similar  properties than (\ref{costOptPcenterCont::eq}) and (\ref{costOptPcenterDiscr::eq}) can be proven to achieve the same complexity of cluster costs.
Lemma \ref{inclusionLemma} 
 has also its proof considering the rectilinear p-center, which allows to prove similarly the Proposition  \ref{propOptdPcent}.
 Capacitated  p-center problems can also be considered,
with bounds on the cardinal of the clusters \cite{calik2015p}.
Proposition  \ref{propOptdPcent} can also be proven in this case,
and Algorithm 2 can be adapted considering the cardinal constraints in the DP programming,
and in the backtracking operations of Algorithm 4.
  
\subsection{Towards a parallel implementation}

The time complexity in $O(NK \log^{1+\gamma} N)$ can be enough for large scale computations of 
$k$-$\gamma$-CP2dPF. 
We mention here that Algorithm 2 with the operators specified in 
Algorithms 3 and 4 have also good properties for a parallel implementation in multi or many-core environments.
As mentioned in section 5.3, the construction of the DP matrix can be computed line by line, with $N-1$ independent computations to compute a line from the one before.
This  property offers a straightforward parallel implementation with OpenMP,
and appropriate structures for a parallelization under General Purpose Graphical Processing Units (GPGPU).
The initial sorting algorithm can also be parallelized on GPGPU (\cite{sintorn2008fast}).
The backtracking algorithm is sequential, but with a low complexity, so that a parallelization of this phase is not crucial. Slight improvements can be processed in a parallelization of  dichotomic searches in Algorithm 4.
Furthermore, the post-processing procedures to have well-balanced clusters as described in section 6.2 are also likely to be parallelized.
The parallelization of the DP algorithm is here useful to reduce the computation time.


\subsection{How to choose $K$?}

A crucial point for the real life application of clustering is to select an appropriate value of $K$.
A too small value of $K$ can miss that an instance is well-captured  with $K+1$ representative clusters.
The real-life application seeks for the best compromise between
the minimization of $K$, and the minimization of the dissimilarity among the  clusters.
Similarly with \cite{dupin2019medoids}, properties of DP can be used in such goal.
With the DP algorithm, many couples $\{(k,C_{k,N})\}_k$ are computed,
the optimal k-center values with $k$ clusters.
Having defined a maximal value $K'$, the complexity for computing these points
is in$O(NK' \log^{1+\gamma} N)$
Searching for good values of $k$, the elbow  technique, graph test, gap test as described in \cite{rasson1994gap}, may be applied. 
Algorithm 4 may be applied for many solutions without changing the complexity.

\subsection{Applications to bi-objective meta-heuristics}

%

The initial motivation of this work was to support the decision makers
when a  MOO approach without preference furnishes a large set of non dominated solutions.
In this application, the value of $K$ is small, for  human analyses to give some preferences.
In this paper, the optimality is not required.
Our work applies also for partial PF furnished by population meta-heuristics \cite{talbi2009metaheuristics}.
A posteriori, the complexity  allows to use Algorithm 3  embedded in MOO approaches, similarly with \cite{auger2009investigating}.
Archiving diversified solutions of Pareto sets has an application for the diversification of genetic algorithms to select diversified solutions for cross-over and mutation phases \cite{zio2011clustering,samorani2018clustering},
but also for  swarm particle optimization heuristics \cite{pulido2004using}.
In these applications, clustering has to run quickly. 
The complexity results and the parallelization properties are useful in such application.

\section{Conclusion and perspectives}

This paper examined properties of the  p-center problems
in the special case of a discrete set of non-dominated points 
 in a 2d Euclidian space.
A common characterization of optimal clusters is proven for the discrete and continuous variants of the p-center  problem.
 It allows to solve these problems to optimality with a unified DP algorithm of a polynomial complexity,
whereas both problems are NP-hard without the PF hypothesis in $\RR^2$.
 A  complexity is proven   in $O(K N \log N)$ time for the continuous p-center problem and  in $O(K N \log^2 N)$ time for the discrete p-center problem, with a space complexity in $O(N)$ for both cases.
 Discrete 2-center is also proven solvable in  $O(N \log N)$ time,
 and 1-center problems in $O(N)$ time.
 It improves general results for 1-center, 2-center problems.
 The DP algorithms can also be  parallelized to speed up the computational time.

Research perspectives are to extend the result of this paper to other clustering algorithms.
In these cases, the crucial point is to prove the interval property of Proposition 
\ref{propOptdPcent}.
In the perspectives to extend some results to dimension 3, interesting for the MOO application, clustering a 3d PF will be a NP-hard problem.
The perspectives are only to find specific approximation algorithms for a 3d PF.




\footnotesize
\bibliographystyle{plain} 

\newpage

\normalsize

%
%
%
%

\section*{Appendix A: Proof of the intermediate lemmas}

This section gives the elementary proofs of intermediate lemmas.

 \vskip 0.5cm

\noindent{\textbf{Proof of Lemma \ref{orderDist}}:} 
We note firstly that the equality cases are trivial, so that we can suppose $ i_1<i_2<i_3$
in the following proof. We prove the propriety (\ref{eq1}), the proof of  (\ref{eq2}) is analogous.\\
Let $ i_1<i_2<i_3$. We note $x_{i_1}=(x^1_{i_1},x^2_{i_1})$, $x_{i_2}=(x^1_{i_2},x^2_{i_2})$  and $x_{i_3}=(x^1_{i_3},x^2_{i_3})$ .\\
Proposition \ref{reord} ordering ensures $x^1_{i_1} < x^1_{i_2} < x^1_{i_3}$ and $x^2_{i_1} > x^2_{i_2} > x^2_{i_3}$.\\
$d(x_{i_1},x_{i_2})^2 = {(x^1_{i_1} - x^1_{i_2})^2 + (x^2_{i_1} - x^2_{i_2})^2}$\\
With $x^1_{i_3} - x^1_{i_1} > x^1_{i_2} - x^1_{i_1}>0$, $(x^1_{i_1} - x^1_{i_2})^2 < (x^1_{i_1} - x^1_{i_3})^2$\\
With $x^2_{i_3} - x^2_{i_1} < x^2_{i_2} - x^2_{i_1}<0$, $(x^2_{i_1} - x^2_{i_2})^2 < (x^2_{i_1} - x^2_{i_3})^2$\\
Thus $d(x_{i_1},x_{i_2})^2 < {(x^1_{i_1} - x^1_{i_3})^2 + (x^2_{i_1} - x^2_{i_3})^2} = d(x_{i_1},x_{i_3})^2$. $\square$

 \vskip 0.5cm
\noindent{\textbf{Proof  of Lemma \ref{costOptPcenterCont}}:} This Lemma is a  reformulated in Lemma \ref{costLemma} below.
 \vskip 0.5cm
\noindent{\textbf{Proof of Lemma \ref{calcKcenter}}:} 
Let $y\in P-\{x_{i},x_{i'}\}$, we denote $j\in [\![i,i']\!]$ such that $y=x_j$.
Applying Lemma \ref{orderDist} to $i<j<i'$: 
\begin{equation}\label{eqInter0}
 \forall k  \in [\![i,i']\!], \;\;\; 
\left|\!\left| x_j - x_k \right|\!\right| \leqslant \max \left(\left|\!\left| x_j - x_i \right|\!\right|,\left|\!\left| x_j - x_i \right|\!\right|\right)
\end{equation}

Finally, Lemma \ref{calcKcenter} is proven with:
 \begin{eqnarray*}
 f_{ctr}^{\DD}(P) & = & \min_{y=x_j \in P} \max_{x \in P}  \left|\!\left| x - y \right|\!\right|\\
  &= &\min_{j \in [\![i,i']\!], x_j \in P} 
  \max \left( \max \left(\left|\!\left| x_j - x_i \right|\!\right|,\left|\!\left| x_j - x_i \right|\!\right| \right), \max_{ k  \in [\![i,i']\!]} \left|\!\left| x_j - x_k \right|\!\right| \right) \\
  &= & 
 \min_{j \in [\![i,i']\!], x_j \in P} \max \left(\left|\!\left| x_j - x_i \right|\!\right|,
 \left|\!\left| x_j - x_{i'} \right|\!\right|  \right) \hfill \square
 \end{eqnarray*}

  \vskip 0.5cm
\noindent{\textbf{Proof of Lemma \ref{inclusionLemma}}:} 
Let $i$ (resp $i'$) the minimal index of points of $P$ (resp $P'$).
Let $j$ (resp $j'$) the maximal index of points of $P$ (resp $P'$).

$f_{ctr}^{\CC}(P) \leqslant f_{ctr}^{\CC}(P')$ is trivial using  Lemma \ref{orderDist} and  Lemma \ref{costOptPcenterCont}.

To prove $f_{ctr}^{\DD}(P) \leqslant f_{ctr}^{\DD}(P')$, we use  Lemma \ref{orderDist} and Lemma \ref{calcKcenter}:


 \begin{eqnarray*}
 f_{ctr}^{\DD}(P) & = & \min_{k \in [\![i,j]\!], x_k \in P} \max \left(\left|\!\left| x_k - x_i \right|\!\right|,\left|\!\left| x_j - x_{k} \right|\!\right|\right)\\
  &\leqslant &\min_{k \in [\![i',j']\!], x_k \in P'} \max \left(\left|\!\left| x_k - x_i \right|\!\right|,\left|\!\left| x_j - x_{k} \right|\!\right|\right) \\
  &\leqslant &\min_{k \in [\![i',j']\!], x_k \in P'} \max \left(\left|\!\left| x_k - x_{i'} \right|\!\right|,\left|\!\left| x_{j'} - x_{k} \right|\!\right|\right) = f_{ctr}^{\DD}(P')  \hfill \square
 \end{eqnarray*}
 \vskip 0.5cm

%
%

   \vskip 0.5cm
 
\noindent{\textbf{Proof of Lemma \ref{monotony}}:}
We define $g_{i,i',j}, h_{i,i',j}$ with:\\
 $g_{i,i'}  :j \in  [\![i,i']\!] \longmapsto  \left|\!\left| x_j - x_i \right|\!\right|$ \hskip 0.2cm and \hskip 0.2cm
$h_{i,i'}  :j \in  [\![i,i']\!] \longmapsto  \left|\!\left| x_j - x_{i'} \right|\!\right|$\\
Let $i<i'$. Proposition \ref{orderDist} applied to $i$ and any $j,j+1$ with  $j\geqslant i$ and $j<i'$
assures that  $g$ is strictly decreasing.
Similarly, Proposition \ref{orderDist} applied to $i'$ and any $j,j+1$  ensures that 
$h$ is strictly increasing.\\
Let $A=\{j  \in [\![i,i']\!] | \forall m  \in [\![i,j]\!] g_{i,i'}(m) < h_{i,i'}(m) \}$.
$g_{i,i'}(i) = 0$ and $h_{i,i'}(i) = \left|\!\left| x_{i'} - x_i \right|\!\right|>0$ so that $i \in A$.
$A$ is a non empty and bounded subset of $\NN$, so that $A$ has a maximum. We note  $l=\max A$.
$h_{i,i'}(i') = 0$ and $g_{i,i'}(i') = \left|\!\left| x_{i'} - x_i \right|\!\right|>0$ so that $i' \notin A$ and $l<i'$.\\
Let $j \in [\![i,l-1]\!]$. 
$g_{i,i'}(j) <   g_{i,i'}(j+1)$ and $h_{i,i',j}(j+1) <   h_{i,i'}(j)$ using monotony of $g_{i,i'}$ and $h_{i,i'}$.
$f_{i,i'}(j+1)= \max \left(g_{i,i'}(j+1),h_{i,i'}(j+1)\right) = h_{i,i}(j+1)$ and  $f_{i,i'}(j)= \max(g_{i,i'}(j),h_{i,i'}(j)) = h_{i,i}(j)$ as $j,j+1 \in A$.
Hence, $f_{i,i'}(j+1)= h_{i,i'}(j+1) <  h_{i,i'}(j) =  f_{i,i'}(j)$.
It proves that $f_{i,i'}$ is strictly decreasing in $[\![i,l]\!]$.\\ 
$l+1 \notin A$ and $g_{i,i'}(l+1) > h_{i,i'}(l+1)$ to be coherent with the fact that $l=\max A$.\\
Let $j \in [\![l+1,i'-1]\!]$. 
$j+1 > j \geqslant l+1$ so $g_{i,i'}(j+1) > g_{i,i'}(j) \geqslant g_{i,i'}(l+1) > h_{i,i'}(l+1) \geqslant h_{i,i'}(j) > h_{i,i'}(j+1)$ using monotony of $g_{i,i'}$ and $h_{i,i'}$.\\
%
%
%
It proves that $f_{i,i'}$ is strictly increasing in $[\![l+1,i']\!]$.\\
Lastly, the minimum of $f$ can be reached in $l$ or in $l+1$, depending on the sign of $f_{i,i'}(l+1)-f_{i,i'}(l)$.
If $f_{i,i'}(l+1)=f_{i,i'}(l)$ there are two minimums $l,l+1$.
Otherwise, there exist a unique minimum $l_0 \in \{l,l+1\}$, $f_{i,i'}$ decreasing strictly  before increasing strictly.  $\square$
 
    \vskip 0.5cm
 
 \noindent{\textbf{Proof of Lemma \ref{monotony02}}:}
 We note firstly that the case $k=1$ is implied by the Lemma \ref{inclusionLemma}, so that we can suppose in the following $k \geqslant 2$.
 Let $k  \in [\![2,K]\!]$ and $j  \in [\![2,N]\!]$.
 Let $\CC_1,\dots, \CC_k$ an optimal solution of
 $k$-center clustering among the points $(x_l)_{l  \in [\![1,j]\!]}$, its cost is $C_{j,k}$. We index the clusters such that $x_j \in \CC_k$.
$$\forall k'  \in [\![1,k]\!], f(\CC_{k'}) \leqslant C_{j,k}$$
We consider the clusters $\CC'_1,\dots, \CC'_k =\CC_1,\dots, \CC_{k-1}, \CC_{k}-{x_k}$. It is a partition of  $(x_l)_{l  \in [\![1,j-1]\!]}$, so that
$C_{j-1,k} \leqslant \max_{k'  \in [\![1,k]\!]} f(\CC'_{k'})$.
Lemma \ref{inclusionLemma} assures that $f(\CC'_{k'}) =f(\CC_{k}-{x_k})\leqslant f(\CC_{k})$.
Hence, $C_{j-1,k} \leqslant \max_{k'  \in [\![1,k]\!]} f(\CC'_{k'}) \leqslant C_{j,k} $, the application $g_{i,k}$  is  increasing. $\square$
 
  \vskip 0.5cm
 
 \noindent{\textbf{Proof of Lemma \ref{monotony2}}:}
This lemma is proven noticing that the following applications are monotone:\\
 $j \in  [\![1,i]\!] \longmapsto  f_{\gamma}(\CC_{j,i})$ is decreasing with Lemma \ref{inclusionLemma}, \\
$j \in  [\![1,N]\!] \longmapsto  C_{j,k}$ is increasing for all $k$ with Lemma \ref{monotony02}. $\square$

 \vskip 0.5cm
 \noindent{\textbf{Proof of Lemma \ref{backtrackOrderLemma}}:}
 This lemma is proven by decreasing induction on $k$, starting
 from $k = K-1$.
 The case $k=K-1$  is furnished by the first step of Algorithm 4,
and that  $j \in  [\![1,N]\!] \longmapsto  f_{\gamma}(\CC_{j,N})$ is decreasing with Lemma \ref{inclusionLemma}.
Having for a given $k$, $i'_k \leqslant i_k$, 
$i_{k-1} \leqslant i'_{k-1}$ is implied by lemma \ref{orderDist} and
$d(z_{i_k},z_{i_{k-1}-1})> OPT$.
$\square$
  \vskip 0.5cm
\begin{lemma}\label{costLemma}
Let $P \subset E$ such that $\mbox{card} (P)\geqslant 2$. 
Let $i$ (resp $j$) the minimal (resp maximal) index of points of $P$.
We have the following results:
\begin{eqnarray}
\forall x \in \RR^2-\left\{\frac {x_i + x_j} 2\right\}, \:\:\:  &   \max_{p \in P} \left|\!\left| x - p \right|\!\right| > \frac 1 2 \left|\!\left| x_i - x_j \right|\!\right| \label{eq1Lemma} \\
 &   \max_{p \in P} \left|\!\left| \frac {x_i + x_j} 2 - p \right|\!\right| = \frac 1 2 \left|\!\left| x_i - x_j \right|\!\right| \label{eq2Lemma}
\end{eqnarray}
In other word, the application $x \in \RR^2 \longmapsto  \max_{p \in P} \left|\!\left| x - p \right|\!\right| \in \RR$ as a unique minimum reached for $x = \frac {x_i + x_j} 2$.

\end{lemma}
\noindent{\textbf{Proof}:} We prove (\ref{eq1Lemma}) and (\ref{eq2Lemma}) using specific coordinates to ease the analytic calculations.
Indeed, the analytic calculus of (\ref{distEucl}) is invariant with translations of the origin and rotations of the axes.
We note $\mbox{diam}(P) = \left|\!\left| x_i - x_j \right|\!\right|$.
\vskip 0.2cm
To prove  (\ref{eq1Lemma}), we use the coordinates defining as origin the point  $O = \frac {x_i + x_j} 2$
and rotating the axes such that in the system of coordinates $x_i = (- \frac 1 2 \mbox{diam}(P),0)$  and $x_j = ( \frac 1 2 \mbox{diam}(P),0)$.
Let $x$ a point in $\RR^2$ and $(x_1,x_2)$ its coordinates. 
We minimize $\max_{p \in P} d(x,x_p)$ distinguishing the cases:\\
if $x_1>0$, $d(x,x_i)^2 = (x_1 +  \frac 1 2 \mbox{diam}(P))^2 +x_2^2 \geqslant (x_1 +  \frac 1 2 \mbox{diam}(P))^2 > (\frac 1 2 \mbox{diam}(P))^2$\\
 $\max_{p \in P} d(x,x_p) \geqslant d(x,x_i) >\frac 1 2 \mbox{diam}(P)$\\
if $x_1<0$, $d(x,x_j)^2 = (x_1 -  \frac 1 2 \mbox{diam}(P))^2 +x_2^2 \geqslant (x_1 -  \frac 1 2 \mbox{diam}(P))^2 > (\frac 1 2 \mbox{diam}(P))^2$\\
 $\max_{p \in P} d(x,x_p) \geqslant d(x,x_j) >\frac 1 2 \mbox{diam}(P)$\\
if $x_1=0$ and $x_2 \neq 0$, $d(x,x_i)^2 = (\frac 1 2 \mbox{diam}(P))^2 +x_2^2 \geqslant (x_1 +  \frac 1 2 \mbox{diam}(P))^2 > (\frac 1 2 \mbox{diam}(P))^2$\\
 $\max_{p \in P} d(x,x_p) \geqslant d(x,x_i) >\frac 1 2 \mbox{diam}(P)$.\\
  The three cases allow to reach any point of $\RR^2$ except $x_0 = \frac {x_i + x_j} 2$, proving (\ref{eq1Lemma}).
\vskip 0.2cm
To prove  (\ref{eq2Lemma}), we use the coordinates translating the axes such that in the system of coordinates $x_i = (- \frac 1 2 \mbox{diam}(P);0)$  and $x_j = ( \frac 1 2 \mbox{diam}(P);0)$.
The origin is the ideal point defined by $x_i$ and $x_j$.
 $x_0$ has coordinates $(\frac 1 {2 \sqrt{2}} \mbox{diam}(P),  \frac 1 {2 \sqrt{2}} \mbox{diam}(P))$.\\
  Let $x=(x^1,x^2) \in P$, such that $x\neq x_i$ and $x\neq x_j$. Thus $ x_i \prec  x_i \prec x_j$.
 The Pareto dominance imposes that  $O \leqslant x_1,x_2\leqslant  \frac 1 {\sqrt{2}} \mbox{diam}(P)$.\\
 $d(x,x_0)^2= (x_1 -  \frac 1 {2 \sqrt{2}} \mbox{diam}(P))^2 +  (x_2 -  \frac 1 {2 \sqrt{2}} \mbox{diam}(P))^2$ \\
 $d(x,x_0)^2 \leqslant  (\frac 1 {2 \sqrt{2}} \mbox{diam}(P))^2 + (\frac 1 {2 \sqrt{2}} \mbox{diam}(P))^2
 = 2 \frac 1 8 \mbox{diam}(P))^2$\\
  $d(x,x_0) \leqslant \frac 1 2 \mbox{diam}(P)$, which proves (\ref{eq2Lemma}) as $d(x_0,x_i) = d(x_0,x_j) = \frac 1 2 \mbox{diam}(P)$.  $\square$

\end{document}